\documentclass[twocolumn]{article}
\usepackage[utf8]{inputenc}
\usepackage{amsmath}
\usepackage{amsfonts}
\usepackage{amssymb}
\usepackage{graphicx}
\usepackage{authblk}
\usepackage{physics}
\usepackage{caption}
\usepackage{geometry}
\usepackage[switch]{lineno}

\def\ba{$^{138}$Ba$^{+}$ }
\def\ban{Ba$^{+}$ }
\def\rb{$^{87}$Rb }

\begin{document}
\title{Demonstration of Slow Light in a Rubidium Vapour Using Single Photons from a Trapped Ion}
\author[1]{J. D. Siverns}
\author[1]{J. Hannegan}
\author[1,2]{Q. Quraishi}
\affil[1]{Joint Quantum Institute, IREAP and Department of Physics, University of Maryland, College Park, MD 20742}
\affil[2]{Army Research Laboratory, Adelphi, MD 20783}
\maketitle
\textbf{Practical implementation of quantum networks are likely to interface different types of quantum systems. When photonic interconnects link the systems together, they must preserve the quantum properties of the photon. These light-matter interfaces may be used as necessary communication tools, such as to synchronise photon arrival times for entanglement distribution. Trapped ions are strong candidates for communication nodes owing to their long qubit life time \cite{langer2005} and high fidelity ion-photon entanglement \cite{stute2012}, whilst neutral atoms are versatile quantum systems, useful as memories \cite{Duan2001, Pan2018, specht2011}, for photon storage \cite{katz2018} or tunable photon delay via slow light \cite{Camacho06,Camacho07}. Development of these two quantum technologies has largely proceeded in separate tracks, partly due to their disparate wavelengths of operation, but combining these platforms offers a compelling hybrid quantum system for use in quantum networking and distributed quantum computing. Here, we demonstrate the first interaction of photons emitted from a trapped ion, with neutral atoms by implementing slow light in a warm atomic vapour.  We use two hyperfine absorption resonances in a warm \rb vapour to provide a slow light medium in which a single photon from a trapped Ba$^+$ ion is delayed by up to 13.5$\pm$0.5 ns. The delay is tunable and  preserves the temporal properties of the photons. This result showcases a hybrid interface, useful for linking different quantum systems together, or as a synchronisation tool for the arrival times of photons - an essential tool for future quantum networks.}

Establishing scalable quantum networks requires integration of disparate quantum components. Significant investment in design, control and development of trapped ion and neutral atom quantum technologies has yielded remarkable progress in quantum networking \cite{hucul2015,Kimble08,kuhn2002}, computing \cite{figgatt2017,haffner2008}, metrology \cite{Wineland2017,Ye2017} and simulation \cite{blatt2012,bloch2012,zhang2017,bernien2017}. Neutral atom vapours and magneto-optical trapped atoms are commonly used as slow-light media \cite{Camacho06,Camacho07,khurgin2010} for either pulses of light or, as in this work, single photons. A hybrid system that combines highly desirable features of different components, such as high-fidelity trapped ion nodes with robust neutral atom quantum storage, would realise a viable quantum network tool. 

Photonic linking of quantum systems to form a hybrid platform has been shown using single atoms \cite{meyer2015,lettner2011}, Bose-Einstein condensates \cite{lettner2011}, solid-state systems \cite{akopian2011}, atomic vapors \cite{Siyushev2014} and atomic ensembles \cite{Zhang2011,maring2017}, although never been trapped ions and neutral atoms. Connecting trapped ions and neutral atoms requires overcoming the disparate wavelengths of the photons emitted by each system, in our case 493 nm for Ba$^+$ ions and 780 nm for neutral Rb. Quantum frequency conversion (QFC) makes it possible to convert a photon's frequency to another frequency, say, to the telecommunication range \cite{Bock2018,Walker18,ikuta2018} or to a wavelength compatible with otherwise incompatible quantum memories \cite{Siverns2018}, while preserving its quantum properties \cite{kumar1990,Bock2018}. Our approach combines QFC and a strong light-mater interaction to form a hybrid trapped ion and neutral atom interface.

\begin{figure*}[t]
	\centering
	\includegraphics[width=\textwidth]{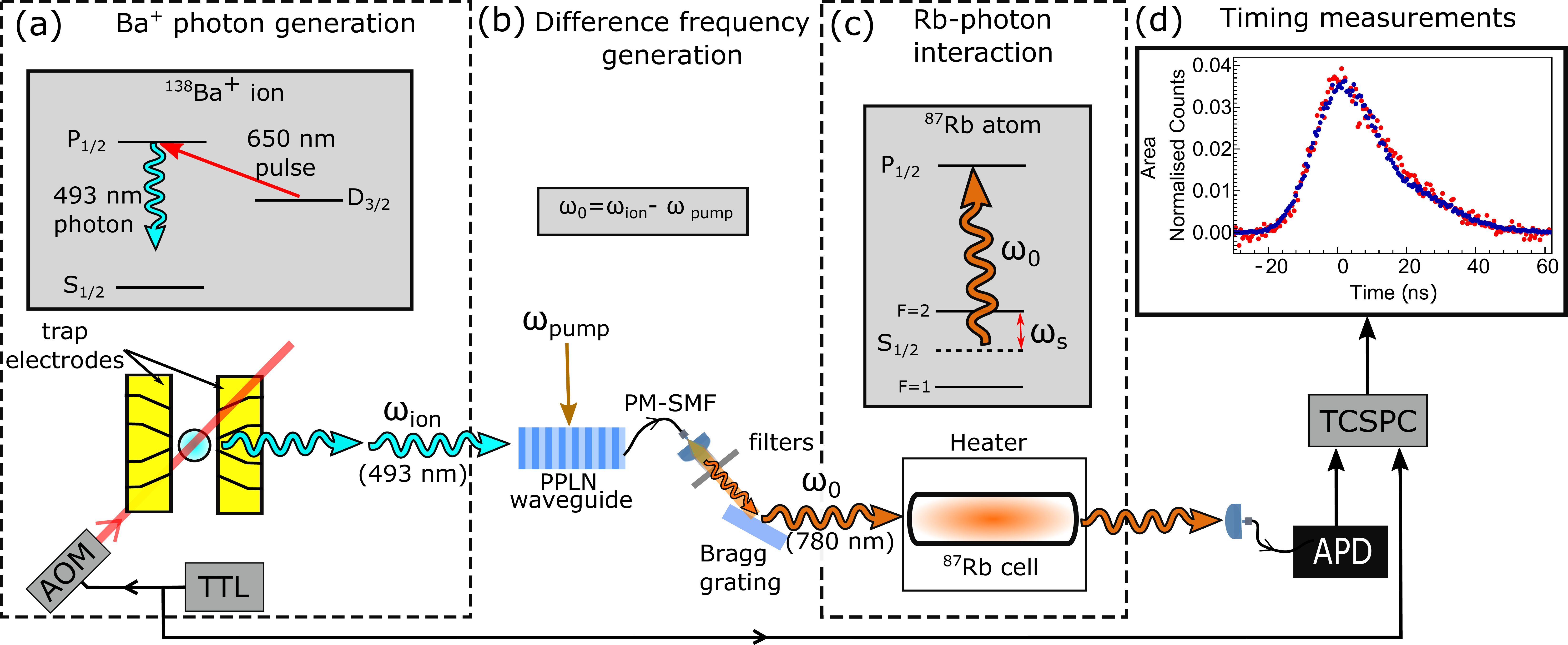}
	\caption{\textbf{Experimental schematic of photon production from a \ba ion, quantum frequency conversion and photonic slowing in a warm neutral \rb vapour} (a) The energy levels of \ba and schematic showing the ion confined in a segmented blade trap. A TTL pulse activated acusto-optic modulator (AOM) controls 650 nm excitation light. (b) The QFC set-up including a periodically-poled-lithium niobate (PPLN) waveguide. Converted light, $\omega_{0}$, is at the difference frequency between ion photons at $\omega_{\text{ion}}$ and pump photons at $\omega_{\text{pump}}$. A series of filters and a Bragg grating filter out pump light and unconverted 493 nm light and reduce the amount of anti-Stokes noise. (c) A \rb energy level diagram and a vapour cell housed inside a heater through which converted single photons pass. (d) Photons are detected on an avalanche photo-diode (APD) and a time-correlated single photon counter (TCSPC) collects the arrival time of the photons with respect to the TTL sent to the AOM. As an example, single photon temporal shapes at 493 nm (blue circles) and frequency converted photons after passing through cell (red circles), both at room temperature.}
	\label{Fig:setup}
\end{figure*}

\begin{figure}
	\centering
	\includegraphics[width=1\columnwidth]{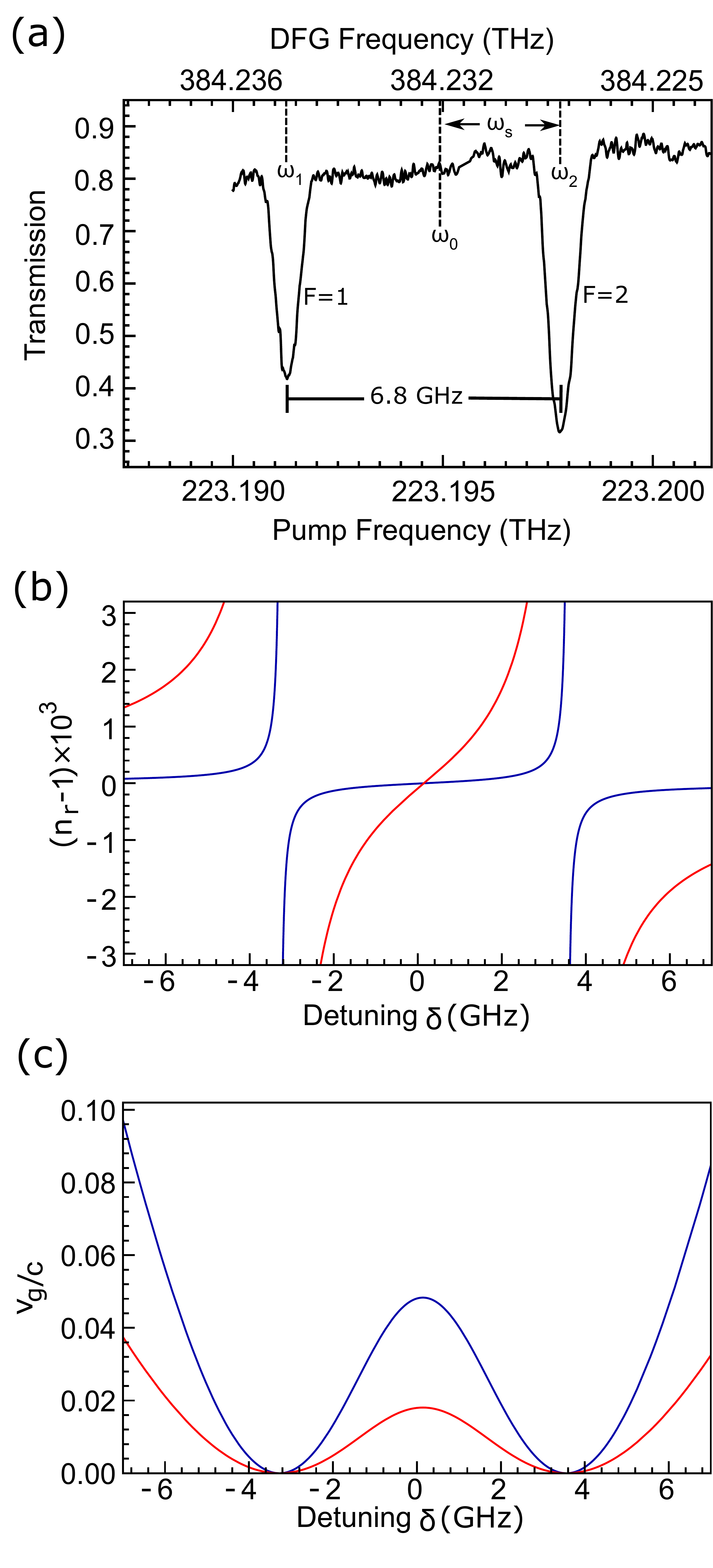}
	\caption{\textbf{Absorption, refractive index and group velocity within a warm $^{87}$Rb vapour} (a) Absorption profile of the $^{87}$Rb $D_2$ line using 780 nm obtained via QFC from 493 nm laser light with the cell at room temperature. The pump laser's mode-hop free tuning range limits the frequency tuning range. The refractive index, (b), and group velocity, (c), in the vicinity of the two absorption peaks as a function of detuning from peak transmission, $\delta$, at 373 K (blue) and 423 K (red).}
	\label{Fig:87Rb}
\end{figure}

Figure~\ref{Fig:setup} shows our experimental set-up used to frequency convert and slow single photons from an ion in a warm atomic vapour cell. Slow light in atomic vapours requires a medium which possesses a low group velocity and can be achieved using photons with a frequency between two absorption resonances of a medium \cite{khurgin2010}. The two absorption resonances may be interrogated via electromagnetically induced transparency (EIT) or far-off resonantly as shown in Fig.~\ref{Fig:87Rb}(a). In this work we use the two D$_{2}$ absorption resonances created by the hyperfine ground state splitting in $^{87}$Rb as described in \cite{Camacho06,Camacho07}. This method requires a less complex experimental set-up as compared with EIT, as only single photons at the correct frequency are required to achieve slowing with no additional lasers required. For this case, the complex index of refraction (Fig.~\ref{Fig:87Rb}(b)) is given by

\begin{equation}
	n(\delta)=1-A\left(\frac{g_1}{\delta + \Delta_+ + i\gamma/2} + \frac{g_2}{\delta - \Delta_- + i\gamma/2}\right)
\label{Eqn:Index}
\end{equation}

\noindent
where $g_1$ and $g_2$ are the the different relative strengths of the two resonances (7/16 and 9/16 respectively for $^{87}$Rb), $\gamma$ is the homogeneous linewidth, $\delta$ is the detuning from peak transmission and $\Delta_{\pm} = \omega_{s} \pm \Delta$ with $ \omega_{s} = (\omega_2 - \omega_1)/2$ and $\Delta = \omega_{s}((g_1^{1/3}-g_2^{1/3})/(g_1^{1/3}+g_2^{1/3}))$. The frequencies of the two individual absorption peaks are given as $\omega_1$ and $\omega_2$. The total strength of the resonance, $A$, in a vapour cell is a function of the atomic number density, $N$, and is given by

\begin{equation}
	A = \frac{N\abs{\mu}^2}{2\epsilon_0\hbar(g_1+g_2)}
\label{Eqn:Strength}
\end{equation}

\noindent
where $\mu$ is the effective far-detuned dipole moment, $\epsilon_0$ is the vacuum permittivity and $\hbar$ is the reduced Plank constant. Using the real part of equation \ref{Eqn:Index}, $n_{r}(\delta)$, it is possible to derive the group velocity,

\begin{equation}
	v_g(\delta) = \left(\frac{\omega_0}{c}\frac{d n_{r}(\delta)}{d\omega}\right)^{-1}
\label{Eqn:GroupVelocity}
\end{equation}

\noindent
where $c$ is the speed of light in vacuum and $\omega_0$ is the frequency from the mid-point between the two absorption resonances to the excited $P_{1/2}$ level and, in this work, the frequency of the converted Ba$^+$ ion photons. From equation \ref{Eqn:GroupVelocity}, in the normal dispersion regime, one can see that a photon with a frequency at the midpoint of two absorption resonances, $\delta = 0$, will have a significantly reduced group velocity while simultaneously experiencing a maximum in transmission (Fig.~\ref{Fig:87Rb}(a) and (c)). It is also clear that if $ N$ in equation \ref{Eqn:Strength} is increased, the group velocity will be reduced. To tune the photon delay, we heat the vapour cell, thereby increasing $N$.

In this work, the source of single photons is a $^{138}$Ba$^{+}$ ion, trapped by segmented blades housed in an ultra-high vacuum chamber (Fig.~\ref{Fig:setup}(a)). To produce single 493 nm photons, the ion is first prepared in the long-lived D$_{3/2}$ level using 493 nm light incident on the ion for $\approx$ 1 $\mu s$. The ion is then excited into the P$_{1/2}$ level using 650 nm light pulsed on for $\approx$ 20 ns using a TTL controlled acousto-optic modulator (AOM) (IntraAction: ATM 80-A1 \cite{disclaimer}). This sequence, interspersed with Doppler cooling with both 493 nm and 650 nm light, is repeated at a rate of $\approx$ 385 kHz [See Methods]. All the optical beams used to interrogate the ion consist of all polarisations. Dead-time is inserted between preparation into the D$_{3/2}$ level and the photon extraction pulses, to ensure that Doppler cooling and state preparation light is not collected in the window of single photon production. This scheme ensures only one 493 nm photon is emitted in the measurement window, as once this color photon is emitted, the ion remains in the ground state with no excitation field present.

The emitted photons are collected using a $\approx$ 0.4 NA lens and sent into a single-mode fibre with a coupling efficiency of $\approx$ 35 $\%$. The fibre-coupled 493 nm photons are then sent to the QFC set-up (Fig.~\ref{Fig:setup}(b)). In the QFC set-up, the ion's photons, frequency $\omega_{ion}$, are combined with a pump laser, frequency $\omega_{pump}$, near 1343 nm and are both free-space coupled into a PPLN waveguide where difference frequency generation (DFG \cite{kumar1990}) occurs due to a $\chi^{(2)}$ non-linearity. This results in photons produced at $\omega_{0}$ (780 nm) $= \omega_{ion} - \omega_{pump} $ \cite{Siverns:17,kumar1990}. By appropriate frequency tuning of the pump laser, we can produce produce 780 nm photons \cite{Siverns2018} with a frequency that is between the two optical absorption resonances of the $^{87}$Rb $D_2$ line, $\omega_0$, (Fig.~\ref{Fig:87Rb}(a)), for implementing slow light.

After the conversion process, the 780 nm photons are passed through a series of optical interference filters to remove pump light and unconverted 493 nm photons (two each of Semrock: LL01-780-25 and FF01-1326/SP-25 \cite{disclaimer}) and with these filters in place, no detectable pump or 493 nm photon counts above the APD dark noise were measured at the pump powers used in this work. Additionally, a Bragg grating with a spectral bandwidth of 0.15 nm and $\approx$ 90$\%$ efficiency was used to filter anti-Stokes noise created by the high intensity pump light \cite{Siverns2018}. With this multi-stage filtering approach it is possible to achieve an end-to-end conversion efficiency of $\approx$ 17$\%$ with a signal to noise ratio (SNR) of $\approx$ 9.8. Once filtered, the photons are sent through a 75 mm long heated glass cell filled with enriched $^{87}$Rb (Triad Technology: TT-RB87-25X-75-P-CAL2O3 \cite{disclaimer}) and then detected on an avalanche photo-detector (APD) (Perkin Elmer: SPCM-AQR-15 \cite{disclaimer}). The inside of the cell is coated with aluminium oxide to reduce rubidium diffusion onto the glass when the cell is at elevated temperatures. Passing photons through the rubidium cell at room temperature causes absorption and scattering of some of the photons and decreases the SNR to $\approx$ 6. The arrival time of the photons at the APD with respect to a 650 nm excitation AOM TTL pulse (Fig.~\ref{Fig:setup}(a)) are measured using a time-correlated single photon counter (TCSPC) with a resolution of 512 ps (PicoQuant: PicoHarp 300 \cite{disclaimer}). Fig.~\ref{Fig:setup}(d) shows the area normalised temporal shape of photons emitted directly from the Ba$^+$ ion at 493 nm and separately after undergoing frequency conversion to a frequency half-way between the two $D_2$ absorption resonances after passing through the $^{87}$Rb vapour cell at room temperature. The arrival time histograms of these two photons clearly shows the preservation of the temporal shape even after QFC and traversal of the cell at room temperature.

\begin{figure}[htbp]
	\centering
	\includegraphics[width=1\columnwidth]{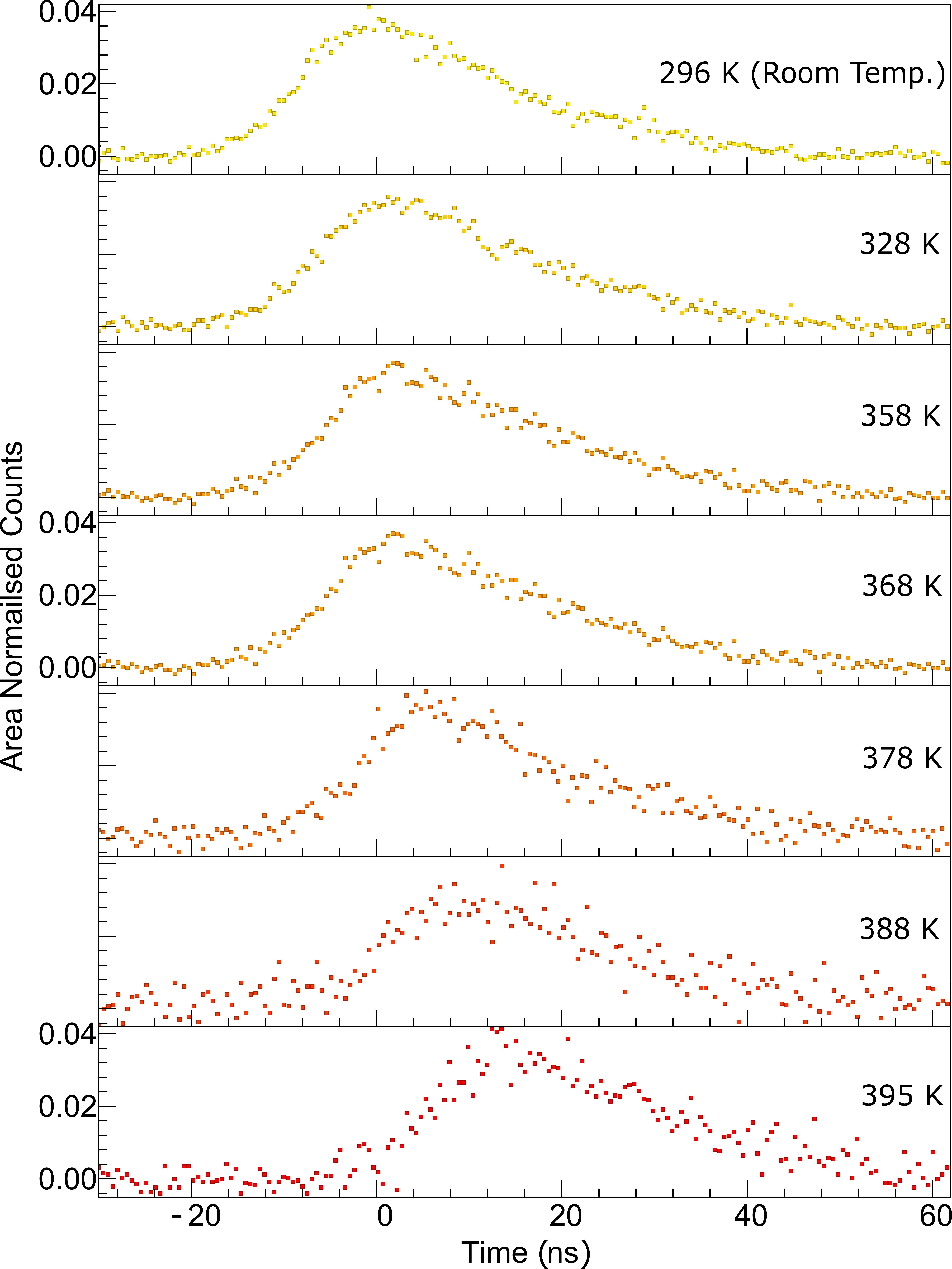}
	\caption{\textbf{Area normalised histogram data of photon temporal shapes at the APD} Temporal shapes of frequency converted photons which have passed through a warm \rb vapour cell. The \rb vapour cell temperature is set at the value indicated near each trace.}
	\label{Fig:slowPhotons}
\end{figure}

\begin{figure}[htbp]
	\centering
	\includegraphics[width=1\columnwidth]{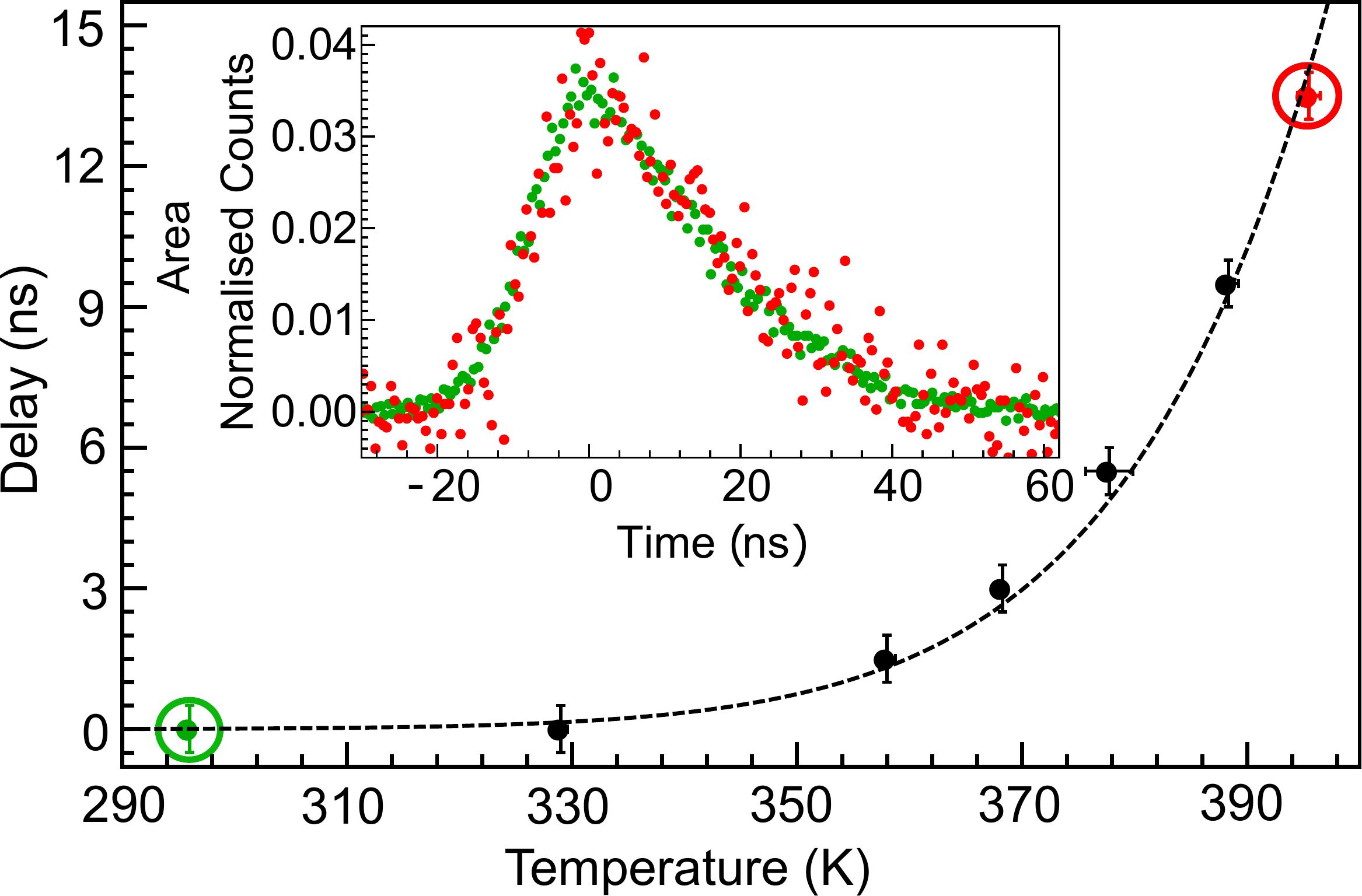}
   	\caption{\textbf{Photon delay as a function of cell temperature} Delay of the frequency converted \ban ion photons after passing through a \rb vapour cell as a function of the cell's temperature. The dashed theory curve is a scaled version of equation \ref{Eqn:GroupVelocity} to account for $N$. The temperature and delay error bars are due to temperature fluctuations over the course of the experiment and the bin width of the histogram photon arrival time data respectively. \textbf{Inset:} Overlap of temporal shapes of photons transmitted through a 296 K room temperature cell (green circles) and a 395 K cell (red circles). The relative delay between the two traces has been removed to allow for comparison.}
	\label{Fig:delay}
\end{figure}

By increasing the temperature of the vapour cell the density of rubidium atoms, $N$, may be increased, leading to a larger change of refractive index and a lower group velocity, as described by equation \ref{Eqn:Index} and \ref{Eqn:GroupVelocity} and shown in Fig.~\ref{Eqn:Index}(b) and (c), respectively. The cell was heated to temperatures ranging from 296 K (room temperature) to 395 K to demonstrate tunable single photon delays of up to 13.5 $\pm$ 0.5 ns (Fig.~\ref{Fig:slowPhotons}). Due to an increase in absorption as the temperature and atomic density of the vapour cell increases, the SNR at the APD decreases from $\approx$ 6 at 296 K to $\approx$ 1 at 395 K. Although there is a lower SNR value for the higher temperature settings, the photon delays are clearly visible and the pulse retains its initial temporal profile.  In Fig.~\ref{Fig:delay}, we plot the photon delay as a function of cell temperature and observe good agreement with the theory curve derived from equation \ref{Eqn:GroupVelocity} where a scaling done is to match the atomic number density, $N$. Both the photons emitted by the Ba$^+$ ion (linewidth $\approx 2\pi\times 15$ MHz) and the drift of the pump laser ($\approx$ 10 MHz) are significantly narrower than the splitting between the $^{87}$Rb $D_2$ absorption peaks ($\approx 6.8$ GHz). This means that all of the converted photons sent through the cell will experience the same delay, resulting in negligible fractional broadening of the photons temporal width irrespective of the cell temperature. The inset of Fig.~\ref{Fig:delay} shows the temporal shape of a single photon passed through the vapour cell at room temperature (green circles) and at 395 K (red circles). This is in contrast to other quantum memories with larger linewidths such as quantum dots, whose photons experience a temporal dispersion and only part of the emitted photons are delayed \cite{akopian2011}. Although the delay is only $\approx$ 0.5 times the temporal width of the photons produced by the Ba$^+$ ion in this experiment, additional delays \cite{Camacho07,kocharovskaya2001} and improved transmission are possible by increasing the non-linear refractive index in the vapour by extending this work using methods such as EIT \cite{katz2018}. 

In summary, we have demonstrated the first interaction of photons emitted from a trapped ion with a neutral atom system by slowing frequency converted ion photons in a warm rubidium vapour cell. Tunable delays of up to 13.5 $\pm$ 0.5 ns were observed with negligible temporal dispersion of the photons, making this system ideal for use as a device for synchronization of remote quantum nodes in a hybrid quantum network. This approach offers a path towards photonic quantum gates between remotely situated ions and neutral atoms, especially given that these two systems can emit photons of comparable temporal profile. Also, this work paves the way for future quantum-state transfer between ions and neutral atoms, experiments such as ion-neutral atom photonic entanglement distribution and photonic storage of flying qubits emitted from trapped ions using existing highly developed neutral atom technologies.
\section*{Acknowledgements}
We would like to thank Alexander Craddock, Dalia Ornelas and Steve Rolston for the use of their Bragg grating and advice in integrating it into our experiment.
\section*{Author Contributions}
J.D.S., J.H. and Q.Q. all contributed to the design, construction, data collection and analysis of the experiment. All authors contributed to the writing of the manuscript. 
\bibliographystyle{unsrt}
\bibliography{slow}

\begin{thebibliography}{10}
\bibitem{disclaimer}
This information is given for technical purposes and does not represent an
  endorsement on the part of the US Government. Other equivalent or better
  options be may available.
\end{thebibliography}
\section*{Methods}

\textbf{Quantum Frequency Conversion from 493 nm to 780 nm}

\begin{figure}[ht]
	\centering
	\includegraphics[width=1\columnwidth]{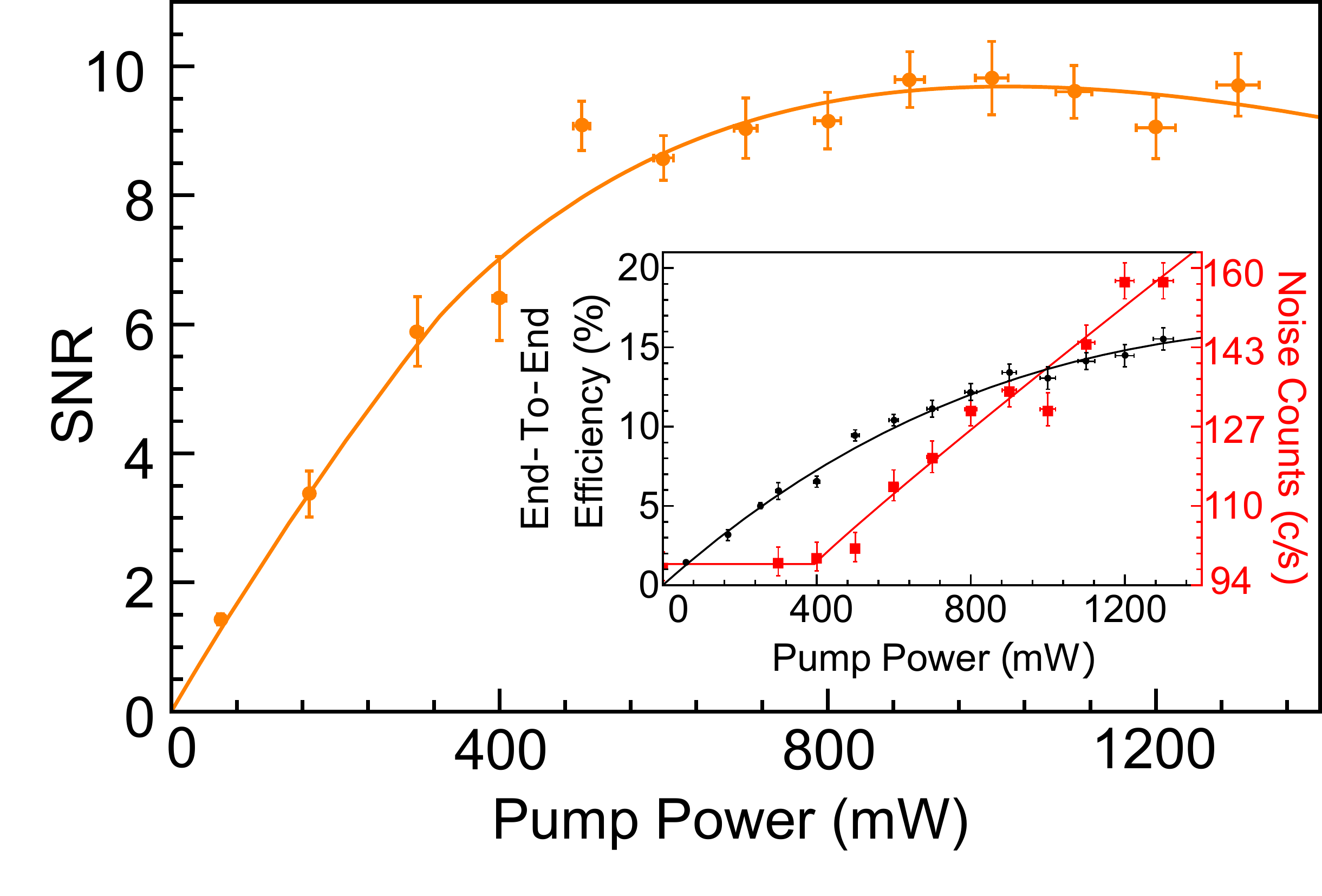}
	\caption{\textbf{Extended Data Figure 1: SNR and DFG curve} SNR measured after filtering of the frequency converted ion signal. The orange curve is the expected SNR given the measured conversion efficiencies and noise at each pump power. \textbf{Inset:} Measured conversion efficiency (black) and measured noise counts (red) on the APD as a function of pump power. The black curve is a theoretical fit to the efficiency data, and the red curve is an empirical fit to the noise.}
    \label{Fig:SNRDFG}
\end{figure}

The conversion efficiency of the PPLN device is a function of the pump power coupled into the waveguide. However, while operating at the maximum conversion efficiency increases the number of 780 nm photons produced, the larger pump power increases the number of anti-Stokes noise photons \cite{pelc2011,Siverns2018}. It is, therefore, critical to maximise the signal-to-noise ratio (SNR) of the converted light and not necessarily the total amount of converted light. Extended Data Fig.\ref{Fig:SNRDFG} shows the SNR achieved in the experiments described in this work. The inset shows both the end-to-end efficiency (defined as the percentage of 493 nm photons entering the QFC set-up that are converted to 780 nm) and the noise counts produced as a function of pump power. For all of these curves, the converted light was sent through optical filters, as described in the main text of the paper. The maximum achieved SNR and end-to-end efficiency were measured to be 9.8$\pm$0.6 and 17.7$\pm$0.7$\%$ respectively. The addition of the narrow bandwidth (0.15 nm) Bragg-grating has allowed an increase in SNR of around a factor of five from our previous work \cite{Siverns2018}. This was achieved without the Rb cell present in the 780 nm beam path. When the room temperature cell is placed on the beam path a 30$\%$ drop in signal is observed due to reflections off of the cell windows and absorption from the Rb atoms. The temperature of the Rb cell is measured both on the cell wall and on a cold finger protruding from the cell. The temperature of this cold finger (the coldest part of the cell) controls the density of atoms in the cell, and this is the reported temperature stated in Fig.~\ref{Fig:slowPhotons} and \ref{Fig:delay}.

\begin{figure}[htbp]
	\centering
	\includegraphics[width=1\columnwidth]{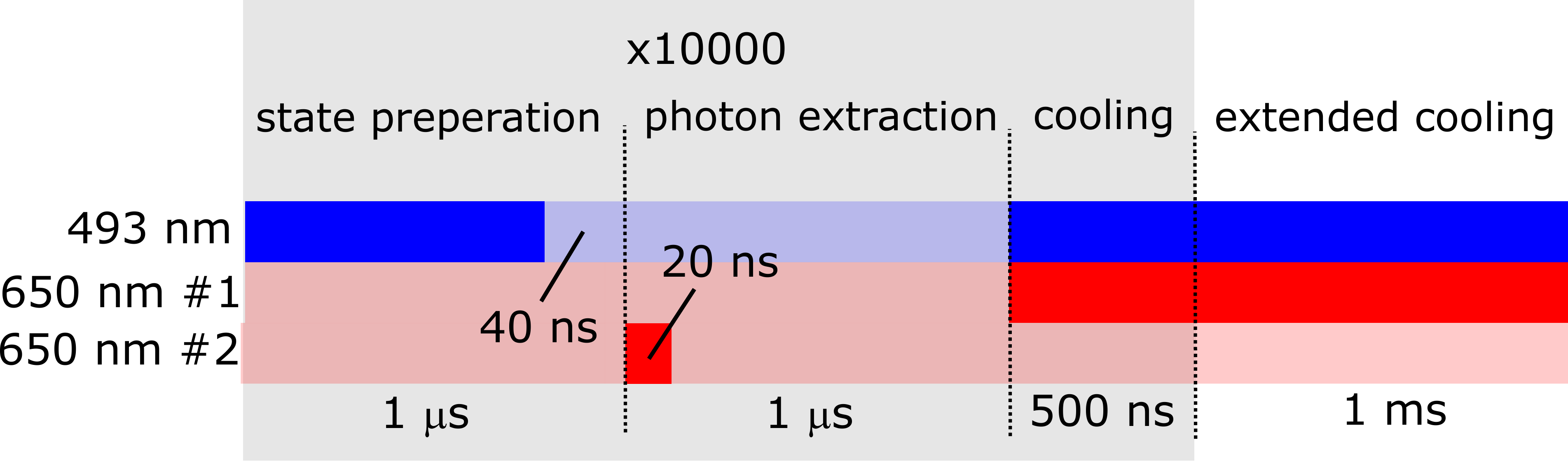}
	\caption{\textbf{Extended Data Figure 2: Experimental pulse sequence.} The experimental pulse sequence showing the timing of the laser beams used to state prepare, extract a single photon and Doppler cool the ion. The parts of the sequence contained in the grey shading are repeated 10,000 times before an extended cooling cycle is performed.}
    \label{Fig:Pulse}
\end{figure}

\vspace{5mm}
\noindent \textbf{Pulse Sequence for Generation of Single Photons from $^{138}$Ba$^+$}\\
To generate single photons, the pulse sequence shown in Extended Data Fig. \ref{Fig:Pulse} is used. First, the ion is prepared into the D$_{3/2}$ state by illuminating the ion with 493 nm light for 1 $\mu$s. Then, a 20 ns 650 nm pulse of light excites the ion into the P$_{1/2}$ state, from which the ion decays into either the S$_{1/2}$ state (with $\approx$ 75\% probability) or back into the D$_{3/2}$ state (with $\approx$ 25\% probability). If the former occurs, a 493 nm photon is emitted and may be collected (with 35\% efficiency) and sent to the frequency conversion set-up. Otherwise, a 650 nm photon is emitted and not detected. After a delay of 940 ns, a short 500 ns cycle of Doppler-cooling is performed before we reinitialize the ion's state into D$_{3/2}$. Importantly, using this scheme ensures that only one 493 nm photon can ever be produced \cite{Siverns:17}. This process is repeated 10,000 times before both lasers are turned on to Doppler cool the ion for 1 ms to prevent excessive heating of the ion. Between preparation and extraction pulses, all beams are turned off for 40 ns to prevent accidental detection of 493 nm leakage light originating from the preparation beam rather than the ion. 

\end{document}